\documentclass[useAMS,usenatbib,a4paper,referee]{mn2e}
\usepackage{graphicx}
\usepackage{amsmath}
\usepackage{txfonts}
\usepackage{mathrsfs}

\newcommand{\be}{\begin{equation}}
\newcommand{\ee}{\end{equation}}
\newcommand{\bea}{\begin{eqnarray}}
\newcommand{\eea}{\end{eqnarray}}

\title[The $R_h=ct$ Universe]{The $R_h=ct$ Universe}
\author[F. Melia and A.S.H. Shevchuk]{F. Melia$^{1}$\thanks{Sir 
Thomas Lyle Fellow and Miegunyah Fellow. E-mail: melia@as.arizona.edu}
and A.S.H. Shevchuk$^{2}$\thanks{E-mail: ashevchuk@as.arizona.edu}\\
$^{1}$Department of Physics, The Applied Math Program, and Department of Astronomy, 
The University of Arizona, AZ 85721, USA\\
$^{2}$Department of Astronomy, The University of Arizona, AZ 85721, USA}

\begin{document}

\date{}

\pagerange{\pageref{firstpage}--\pageref{lastpage}} \pubyear{2010}

\maketitle

\label{firstpage}

\begin{abstract}
The backbone of standard cosmology is the Friedmann-Robertson-Walker
solution to Einstein's equations of general relativity (GR). In recent
years, observations have largely confirmed many of the properties
of this model, which is based on a partitioning of the universe's
energy density into three primary constituents: matter, radiation,
and a hypothesized dark energy which, in $\Lambda$CDM, is assumed
to be a cosmological constant $\Lambda$. Yet with this progress,
several unpalatable coincidences (perhaps even inconsistencies) have
emerged along with the successful confirmation of expected features.
One of these is the observed equality of our gravitational horizon
$R_{\rm h}(t_0)$ with the distance $ct_0$ light has traveled since
the big bang, in terms of the current age $t_0$ of the universe. This
equality is very peculiar because it need not have occurred at all and,
if it did, should only have happened once (right now) in the context
of $\Lambda$CDM. In this paper, we propose an explantion for why
this equality may actually be required by GR, through the
application of Birkhoff's theorem and the Weyl postulate,
at least in the case of a flat spacetime. If this proposal is
correct, $R_{\rm h}(t)$ should be equal to $ct$ for all cosmic
time $t$, not just its present value $t_0$. Therefore models such as
$\Lambda$CDM would be incomplete because they ascribe
the cosmic expansion to variable conditions not consistent with
this relativistic constraint. We show that this may be the reason
why the observed galaxy correlation function is not consistent with
the predictions of the standard model. We suggest that an $R_{\rm h}=ct$
universe is easily distinguishable from all other models at large
redshift (i.e., in the early universe), where the latter all
predict a rapid deceleration.
\end{abstract}


\section{Introduction}
The standard model of cosmology, $\Lambda$CDM, is today confronted with several
inconsistencies and unpalatable coincidences, even though it arguably represents
the most successful attempt at accounting for the cosmological observations.
Many have written extensively on this subject, including, e.g., Spergel et al.
(2003), and Tegmark et al. (2004). For example, $\Lambda$CDM has been used with
measurements of the cosmic microwave background (CMB) radiation to infer that
the universe is flat, so its energy density $\rho$ is at (or very near) its
``critical" value
\begin{equation}
\rho_{\rm c}\equiv 3c^2H^2/8\pi G\;,
\end{equation}
where $H$ is the Hubble constant and the other symbols have their usual
meanings. Yet among the many
peculiarities of the standard model is the inference that the density
$\rho_{\rm de}$ of dark energy must itself be of order $\rho_{\rm c}$.
Worse, no reasonable explanation has yet been offered as to why such a
fixed, universal density ought to exist at this scale. It is well known
that if $\Lambda$ is associated with the energy of the vacuum in quantum
theory, it should have a scale representative of phase transitions in the
early universe---120 orders of magnitude greater than $\rho_{\rm c}$.

The most recent---and perhaps most disturbing---coincidence with
$\Lambda$CDM is the apparent equality of our gravitational horizon
$R_{\rm h}(t_0)$ with the distance $ct_0$ light has traveled since the big
bang (in terms of the presumed current age $t_0$ of the universe). This
equality was first identified as a peculiarity of the standard model in
Melia (2003), and has come under greater scrutiny in recent years
(Melia 2007, 2009; Melia \& Abdelqader 2009; van Oirschot et al. 2010;
see also Lima 2007 for a related, though unpublished, work).

The purpose of this paper is to advance a possible explanation for why the
observed equality $R_{\rm h}(t_0)=ct_0$ may in fact not be a coincidence
of any particular model, such as $\Lambda$CDM. Rather, we suggest a reason
why it may be required for all cosmologies, by an application of Birkhoff's
theorem and its corollary, together with the Weyl postulate, to the properties
of the Friedmann-Robertson-Walker spacetime.
More importantly, we show that, at least for flat cosmologies, this
equality may actually be upheld for all cosmic time $t$ which, however,
would not be entirely consistent with $\Lambda$CDM, or any other cosmological
model we know of. We shall see that if our proposal turns out to be correct,
models such as $\Lambda$CDM would then be compelled to fit the data subject
to the constraint $R_{\rm h}(t_0)=ct_0$ today, but would therefore incorrectly
ascribe the universal expansion to variable conditions inconsistent with this
time-independent GR-Weyl constraint in the past. We conclude by suggesting
that an $R_{\rm h}=ct$ universe is unmistakably distinguishable from all
other models through a comparison with standard candles at redshifts extending
beyond the current Type Ia supernova limit at $\sim1.8$, therefore providing
a reliable test of our proposal when compared to other models.

\section[]{The FRW Equations}
Standard cosmology is based on the Friedmann-Robertson-Walker (FRW) metric for a spatially
homogeneous and isotropic three-dimensional space, in which the coordinates expand or
contract as a function of time:
\begin{equation}
ds^2=c^2\,dt^2-a^2(t)[dr^2(1-kr^2)^{-1}+
r^2(d\theta^2+\sin^2\theta\,d\phi^2)]\;.
\end{equation}
The coordinates for this metric have been chosen so that $t$ represents the time measured by a
comoving observer (and is the same everywhere, so it functions as a ``community" time),
$a(t)$ is the expansion factor, and $r$ is an appropriately scaled radial coordinate
in the comoving frame. The geometric factor $k$ is $+1$ for a closed universe, $0$ for a
flat, open universe, or $-1$ for an open universe.

Applying the FRW metric to Einstein's field equations of GR, one obtains
the corresponding FRW differential equations of motion. These are the Friedmann equation,
\begin{equation}
H^2\equiv\left({\dot a\over a}\right)^2={8\pi G\over 3c^2}\rho-{kc^2\over a^2}\;,
\end{equation}
and the ``acceleration" equation,
\begin{equation}
{\ddot a\over a}=-{4\pi G\over 3c^2}(\rho+3p)\;.
\end{equation}
An overdot denotes a derivative with respect to cosmic time $t$, and $\rho$ and $p$
represent the total energy density and total pressure, respectively. A further application
of the FRW metric to the energy conservation equation in GR yields the final equation,
\begin{equation}
\dot\rho=-3H(\rho+p)
\end{equation}
which, however, is not independent of Equations~(3) and (4).

\section{The Birkhoff Theorem and the Observer's Gravitational Horizon}
In comoving coordinates, the proper distance $R(t)$ is measured at constant $t$ and
one can easily see from Equation~(2) that for purely radial paths in a flat
cosmology, $R(t)=a(t)r$. It is sometimes useful to recast Equation~(2) in terms
of $R(t)$ (see Equation~9 below) which can reveal, e.g., the dependence of the metric
coefficients on the observer's gravitational horizon, which we now define.

The Hubble radius is the point at which the universal expansion rate $\dot{R}(t)=
\dot{a}(t)r$ equals the speed of light $c$. But though this radius is well known,
it is rarely recognized as just a manifestation of the gravitational radius (see
Melia 2007), because every observer experiences zero {\it net} acceleration from 
a surrounding isotropic mass, suggesting that no measure of distance equivalent 
to the Schwarzschild radius is present in cosmology.

But in fact the {\it relative} acceleration between an observer and any other
spacetime point in the cosmos is not zero; it depends on the mass-energy
content between him/herself and that point. This is most easily understood in
the context of Birkhoff's theorem and its corollary (Birkhoff 1923)---a
relativistic generalization of Newton's theory, that the gravitational
field outside a spherically symmetric body is indistinguishable from that
of the same mass concentrated at its center.

What is particularly germane to our discussion here is the
corollary to this theorem, describing the field {\it inside} an empty spherical
cavity at the center of an isotropic distribution. The metric inside such a
cavity is equivalent to the flat-space Minkowski metric $\eta_{\alpha\beta}$,
a situation not unlike that found in electromagnetism, where the electric field
inside a spherical cavity embedded within an otherwise uniform charge
distribution is zero. Not surprisingly, the corollary to Birkhoff's theorem
is itself analogous to another Newtonian result---that the gravitational
field of a spherical shell vanishes inside the shell. So even in the classical
limit, one can argue that the medium exterior to a spherical cavity
may be thought of as a sequence of ever increasing spherical shells, each
of which produces a net zero effect within the cavity.

To understand the emergence of a gravitational radius in cosmology, imagine 
placing an observer at the center of this spherical cavity with proper radius
$R_{\rm cav}$, surrounding him/her by a spherically-symmetric mass with a proper
surface radius $R_{\rm s}<R_{\rm cav}$. The metric in the space between the
mass and the edge of the cavity is given by the Schwarzschild solution, and
the relative acceleration between the observer and $R_{\rm s}$ is simply due
to the mass enclosed within $R_{\rm s}$, which we may write in terms of
the cosmic energy density $\rho(t)$ as
\begin{equation}
M(R_{\rm s})=V_{\rm prop}\,{\rho(t)\over c^2}\;,
\end{equation}
where
\begin{equation}
V_{\rm prop}={4\pi\over 3}R_{\rm s}^3
\end{equation}
is the proper volume.\footnote{To be absolutely clear about this definition,
we emphasize the fact that $V_{\rm prop}$ is the volume within which the
co-moving density of particles remains fixed as the universe expands.}

The criterion we will use to define the gravitational radius $R_{\rm h}$  is
\begin{equation}
R_{\rm h}\equiv {2GM(R_{\rm h})\over c^2}
\end{equation}
(see Melia 2007, Melia \& Abdelqader 2009). As we shall see below,
the FRW equations in principle allow many different kinds of solutions with
their own particular form of the expansion factor $a(t)$. When we impose the
condition in Equation~(8), however, only one of these solutions is permitted.
This unique solution corresponds to the observed equality
$R_{\rm h}(t_0)=ct_0$, which is most easily inferred from the measurement of
$H_0$ in the SHOES project (Riess et al. 2009), refining the value previously
obtained through the Hubble Space Telescope Key Project on the extragalactic
distance scale (Mould et al. 2000). The Hubble constant, $H_0\equiv H(t_0)= 
74.2\pm3.6$ km s$^{-1}$ Mpc$^{-1}$, is now known with unprecedented accuracy.
In the context of $\Lambda$CDM, the density $\rho$ is at (or very near)
its ``critical" value $\rho_{\rm c}$, and with this $H_0$,
$R_{\rm h}(t_0)\approx 13.7$ billion lightyears ($\approx ct_0$).

Equation~(8) explains why the Hubble radius exists in the first place, and
is our proposal for a resolution of the $R_{\rm h}(t_0)=ct_0$ coincidence 
in the standard model. Ironically, though many may be unaware of the 
existence of this radius, de Sitter's own solution to Einstein's equations 
was actually first written in terms of what we now call the proper distance
$R(t)=a(t)r$; a limiting radius equivalent to $R_{\rm h}$ appeared
in his form of the metric (see de Sitter 1917).

It is now well known that de Sitter's spacetime describes a universe
driven by an exponential scale factor $a(t)$. In the more general
case, it is not difficult to show, in terms of the proper radii
$R$ and $R_{\rm h}$, that Equation~(2) transforms to
\begin{equation}
ds^2= \Phi\, c^2dt^2 + 2\left(\frac{R}{R_{\rm h}}\right)c\,dt\,dR  - 
{dR^2}-R^2\,d\Omega^2
\end{equation}
(Melia \& Abdelqader 2009), where the function
\begin{equation}
\Phi\equiv 1-\left(\frac{R}{R_{\rm h}} \right)^2
\end{equation}
signals the dependence of the metric on the proximity of the proper radius
$R$ to the gravitational radius $R_{\rm h}$. We have here assumed a flat
universe with $k=0$, as indicated by the precision measurements of the
CMB radiation (Spergel et al. 2003). The reader will also notice that,
formally, $R_{\rm h}$ functions as the {\it static} limit, since the
interval $ds$ becomes unphysical at any {\it fixed} proper distance
$R$ beyond $R_{\rm h}$. However, there is no such exclusion on the
viability of this metric beyond $R_{\rm h}$ when $\dot{R}\not=0$,
such as we have for sources receding from us with the Hubble
expansion (more on this below).

The impact of Equation~(8) may now be gauged with the use of Equation~(3),
yielding (with $k=0$)
\begin{equation}
R_{\rm h}={c\over H(t)}=c{a\over \dot{a}}
\end{equation}
(see Melia \& Abdelqader 2009). This is in fact also the
definition of the better known Hubble radius, which is therefore simply
another manifestation of the gravitational radius $R_{\rm h}$. Thus,
given what we know about the analogous gravitational radius of a static
spherical mass, it is not surprising that the expansion rate $\dot{R}$
should equal $c$ when $R\rightarrow R_{\rm h}$, just as the speed of
matter falling towards a black hole reaches $c$ at the event horizon.
This may be seen most easily from the definition of $R$ and Equation~(11),
which together give
\begin{equation}
\dot{R}=c{R\over R_{\rm h}}\;,
\end{equation}
and therefore $\dot{R}=c$ when $R=R_{\rm h}$.
Below we analyze the role of $R_{\rm h}$ further and see that, even
though Equation~(9) is quite general as written, the definition of the
gravitational radius in Equation~(8) actually selects out only one specific
FRW solution, which we are proposing as the correct cosmic spacetime.

\section{Consistency with the Weyl Postulate}
As a prelude to our further consideration of $R_{\rm h}$, we
reaffirm the fact that the universe appears to be homogeneous
and isotropic on large scales, meaning that observations made
from our vantage point are representative of the cosmos as viewed
from anywhere else. Known as the Cosmological Principle, the
assumption of homogeneity and isotropy is essential to any attempt
at using what we see here from Earth as a basis for testing
cosmological models.

On large scales, at least, the universe appears to be expanding in an 
orderly manner, with galaxies moving apart from one another (except for
the odd collision or two due to some peculiar motion on top of the
``Hubble flow"). Galactic trajectories on a spacetime diagram
would therefore show world lines forming a funnel-like structure
in which the separation between any two paths is steadily increasing
with time $t$.

Homogeneity and isotropy are consistent with this type of regularity,
and together suggest that the evolution of the universe may be
represented as a time-ordered sequence of three-dimensional
spacelike hypersurfaces, each of which satisfies the
Cosmological Principle. This intuitive picture of regularity
is often expressed formally as the {\it Weyl postulate}, after
the mathematician Hermann Weyl, who did much of the early work
on this subject in the 1920's (see, e.g, Weyl 1923).

The most general line element satisfying the Weyl postulate
and the Cosmological Principle is given by Equation~(2) above,
in which the spatial coordinates $(r,\theta,\phi)$ are constant
from  hypersurface to hypersurface in the expanding flow, while
the temporal behavior of the scale factor $a(t)$ reflects
the dynamics of the expanding cosmos. This metric was first
rigorously derived in the 1930's by Robertson (1935) and
(independently) Walker (1936), using the ideas espoused
earlier by Weyl.

It is therefore clear that any proper distance in this spacetime
is measured on a spacelike hypersurface in the foliated sequence
orthogonal to the non-intersecting geodesics. We have shown in
\S~3 above that the Hubble radius is itself the distance
$R_{\rm h}$. But according to the definition of $R_{\rm h}$ in
terms of $V_{\rm prop}$ in Equation~(8), $R_{\rm h}$ must
itself be a proper distance
\begin{equation}
R_{\rm h}=a(t)r_{\rm h},
\end{equation}
with the property that $r_{\rm h}$ is a constant comoving
coordinate, otherwise $V_{\rm prop}$ would not represent
the volume within which the particle density is constant
in the comoving frame. Comparing Equations~(11) and (13),
we therefore see that
\begin{equation}
r_{\rm h}\equiv {c\over\dot{a}}\;,
\end{equation}
which means that $\dot{a}$ itself must be constant for
consistency with the Weyl postulate. This is the most
important consequence of our definition of $R_{\rm h}$
in Equation~(8).

From Equation~(4), we infer that the acceleration $\ddot{a}$
is zero either for an empty universe (in which $\rho=p=0$) or
one characterized by an equation of state $w=-1/3$ (refer to
the Appendix for some additional insight into why these two
conditions are actually related). And it is trivial to see
from Equation~(11) that
\begin{equation}
R_{\rm h}=ct
\end{equation}
for all cosmic times $t$, not just the current value $t_0$.

Within the framework of our proposal, one may then understand
why today we ``measure" $R_{\rm h}(t_0)$ to be equal to $ct_0$
(within the observational errors), because in a flat universe
($k=0$) consistent with the Weyl postulate and the Cosmological
Principle, these two quantities must always be equal.

\section{Cosmological Models}
Let us now see how this result impacts the standard model of
cosmology. We suppose that
\begin{equation}
\rho=\rho_{\rm m}+\rho_{\rm r}+\rho_{\rm de}
\end{equation}
where, following convention, we designate the matter, radiation, and dark
energy densities, respectively, as $\rho_{\rm m}$, $\rho_{\rm r}$, and
$\rho_{\rm de}$. We will also assume that these energy densities scale
according to $\rho_{\rm m}\propto a^{-3}$, $\rho_{\rm r}\propto a^{-4}$,
and $\rho_{\rm de}\propto f(a)$. (If dark energy is indeed a cosmological
constant $\Lambda$, then $f(a)=$ constant.) Thus, defining
\begin{equation}
\Omega_{\rm m}\equiv {\rho_{\rm m}(t_0)\over\rho_{\rm c}}\;,
\end{equation}
\begin{equation}
\Omega_{\rm r}\equiv {\rho_{\rm r}(t_0)\over\rho_{\rm c}}\;,
\end{equation}
and
\begin{equation}
\Omega_{\rm de}\equiv {\rho_{\rm de}(t_0)\over\rho_{\rm c}}\;,
\end{equation}
with the (flatness) constraint
\begin{equation}
\Omega_{\rm m}+\Omega_{\rm r}+\Omega_{\rm de}=1\;,
\end{equation}
we may rewrite the Friedmann equation as
\begin{equation}
\left({da\over dt}\right)^2 = {H_0}^2\left\{1+
\Omega_{\rm m}\left({1\over a}-1\right)+\Omega_{\rm de}(a^2f-1)\right\}\;.
\end{equation}
We have here normalized the expansion factor so
that $a(t_0)=1$, which we assume throughout this paper.

Introducing the cosmological redshift $z$, where
\begin{equation}
1+z={1\over a(t)}\;,
\end{equation}
we can re-arrange this equation to read
\begin{equation}
{1\over(1+z)^2}{dz\over dt}=- {H_0}\left\{1+
\Omega_{\rm m}\left({1\over a}-1\right)+\Omega_{\rm de}(a^2f-1)\right\}^{1/2}\;,
\end{equation}
so that
\begin{equation}
H_0\int_{t_e}^{t_0} dt=\int_0^{z(t_e)}{dz\over (1+z)^2[1+\Omega_{\rm m}
z-g(z)\Omega_{\rm de}]^{1/2}}\;.
\end{equation}
That is
\begin{equation}
c(t_0-t_e)=R_{\rm h}(t_0)\int_0^{z(t_e)}{dz\over (1+z)^2[1+\Omega_{\rm m}
z-g(z)\Omega_{\rm de}]^{1/2}}\;,
\end{equation}
where we have also defined the function $g(z)\equiv f/(1+z)^2-1$,
and $z(t_e)$ is the redshift of light reaching us at $t_0$, but
emitted at cosmic time $t_e$. In this expression, we have used the
equality $R_{\rm h}=c/H$, which is valid in a flat ($k=0$) cosmology.
Other than this flat condition, Equation~(25) is identical to that obtained
in the concordance model, subject to the density in Equation~(16).

If we now put $t_e\rightarrow 0$ and $z(t_e)\rightarrow \infty$,
then clearly
\begin{equation}
ct_0=R_{\rm h}(t_0)\int_0^\infty{dz\over (1+z)^2[1+\Omega_{\rm m}
z-g(z)\Omega_{\rm de}]^{1/2}}\;.
\end{equation}
Our proposed form of the gravitational (i.e., Hubble) radius
in Equation~(8) leads to the equality $R_{\rm h}(t_0)=ct_0$.
Therefore, any cosmological model consistent with the Weyl
Postulate and the Cosmological Principle must satisfy the condition
\begin{equation}
\int_0^\infty{dz\over (1+z)^2[1+\Omega_{\rm m}
z-g(z)\Omega_{\rm de}]^{1/2}}=1\;.
\end{equation}
Although not immediately obvious, this constraint implies that no matter
what period of deceleration or acceleration the universe may have
experienced in its past, its overall acceleration averaged over the time
$t_0$ must be zero (Melia 2009). We can best see this directly from the
FRW equations, which indicate that
\begin{equation}
\dot{R_{\rm h}}\equiv {dR_{\rm h}\over dt}={3\over 2}
(1+w)c\;,
\end{equation}
where the parameter
\begin{equation}
w\equiv {p\over\rho}
\end{equation}
characterizes the total pressure $p$ in terms of the total energy density
$\rho$. Under the assumption that $R_{\rm h}$ was much smaller in the
distant past than it is today, we can easily integrate this equation to get
\begin{equation}
R_{\rm h}(t_0)={3\over 2}(1+\langle w\rangle)ct_0\;,
\end{equation}
where
\begin{equation}
\langle w\rangle\equiv {1\over t_0}\int_0^{t_0}w(t)\,dt\;.
\end{equation}
Thus, in order for $R_{\rm h}(t_0)$ to equal $ct_0$ (which in turn leads
to Equation~27), we must have $\langle w\rangle=-1/3$, corresponding to
an average acceleration $\langle\ddot{a}\rangle=0$ in Equation~(4).

Any cosmological model that purports to correctly trace the universal
expansion must simultaneously satisfy Equation~(27) and the condition
$\langle w\rangle=-1/3$. In $\Lambda$CDM, for example, dark energy
is considered to be a cosmological constant, so $g(z)=z(2+z)/(1+z)^2$.
In figure~1, we plot the value of the integral in Equation~(27) as a
function of $\Omega_{\rm m}$ for a flat $\Lambda$CDM
cosmology. Not surprisingly, the integral is $1$ when $\Omega_{\rm m}
\approx 0.27$, consistent with the optimized parameters of the
concordance model (see, e.g., Spergel et al. 2003).

\begin{figure}
\center{\includegraphics[scale=0.55,angle=0]{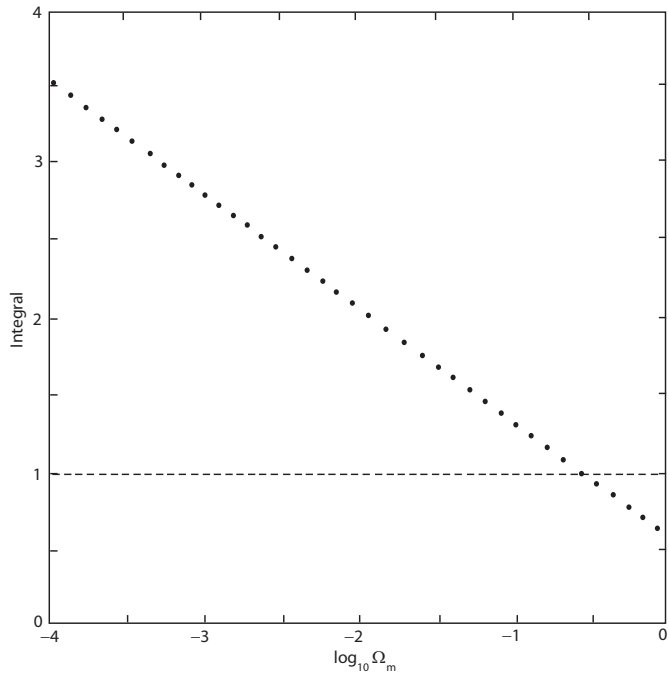}
\caption{The integral in Equation~(27) as a function of $\Omega_{\rm m}$,
assuming a flat cosmology, for the standard model (i.e., $\Lambda$CDM). The
integral equals $1$ when $\Omega_{\rm m}\approx 0.27$ (and $\Omega_{\rm de}
\equiv\Omega_\Lambda \approx 0.73$). It is important to emphasize that this
inferred value of $\Omega_{\rm m}$ comes, not from fits to the cosmological
data using the $\Lambda$CDM decomposition in Equation~(16) but, rather, 
from the imposition of the Weyl postulate expressed through Equation~(27).}}
\end{figure}

Using the same optimized parameters to evaluate the
integral in Equation~(31), we obtain the time-averaged value of $w$
plotted as a function of cosmic time in figure~2. We see that
$\langle w\rangle\approx -1/3$ at $t\approx 1/H_0$, consistent
with the fit shown in figure~1. Clearly, the simplest way to satisfy
both Equation~(27) and the constraint $\langle w\rangle=-1/3$ would be
to have $w=-1/3$ for all cosmic time $t$. But this is not what
happens in $\Lambda$CDM, as one can trivially see from figure~2.
Instead, one must adjust the values of $\Omega_{\rm m}$ and
$\Omega_{\rm de}$ in order to make the integral in Equation~(27)
come out to $1$, which ensures that $\langle w\rangle=-1/3$ today,
but neither $w$ nor $\langle w\rangle$ are equal to $-1/3$ at any
other time. This is far from satisfactory, however, because (as
noted previously by Melia 2009), the time-averaged value of $w$
could then be equal to $-1/3$ only once in the entire history
of the universe, and that would have to happen right now.

\section{The Luminosity Distance}
The distinction between our proposed cosmology with $R_{\rm h}=ct$
(for all $t$, not just $t_0$), and other FRW models with past
epochs of deceleration, is quite pronounced at redshifts larger
than the current limits ($\sim 1.5-2$) of study. This happens
because the application of Birkhoff's theorem, together with
the Weyl postulate and the Cosmological Principle, suggests
that $w=-1/3$ for all $t$, whereas $\langle w\rangle$ in
$\Lambda$CDM changes with cosmic time (see figure~2).

Based on current Type Ia supernova measurements, the use of
$\Lambda$CDM as the standard evolutionary model seems to
provide an adequate fit to the data. This could present a
problem for our proposal because our explanation for the observed
equality $R_{\rm h}(t_0)=ct_0$ would suggest that the $\Lambda$CDM
version of the luminosity distance $d_{\rm L}$ used to fit the
Type Ia supernova data (e.g., the ``gold sample" in Riess et al.
2004) is not correct in a flat spacetime (see also Riess et al.
1998, and Perlmutter et al. 1999). However, the disparity between
this version of $d_{\rm L}$ and that required by a flat cosmology
with $w=-1/3$, increases with redshift, so in principle we
should be able to distinguish between the two by observing
events at sufficiently early times.

\begin{figure}
\center{\includegraphics[scale=0.55,angle=0]{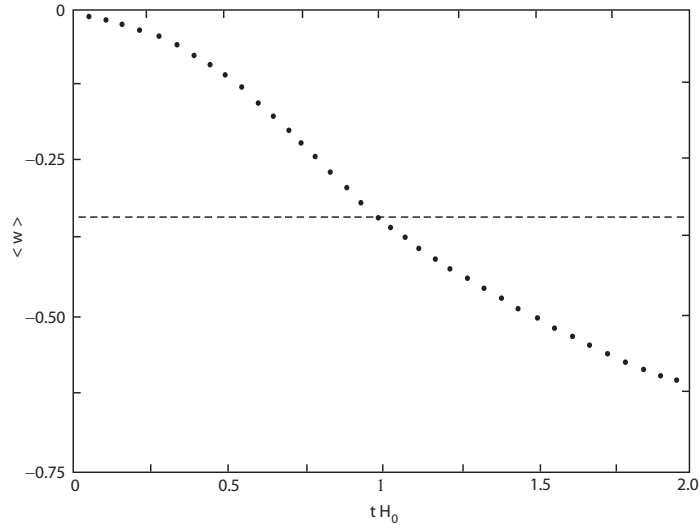}
\caption{Average value of $w$ (Eq.~29) as a function of 
$t_0$, in units of $1/H_0$, for the standard model (i.e., $\Lambda$CDM). 
The dashed line corresponds to $\langle w\rangle=-1/3$.}}
\end{figure}

In $\Lambda$CDM, the luminosity distance is given as
\begin{equation}
d_{\rm L}=(1+z)\,R_{\rm h}(t_0)\int_0^z{du\over
[\Omega_{\rm m}(1+u)^3+\Omega_{\rm de}(1+u)^\alpha]^{1/2}}
\end{equation}
where, strictly speaking, dark energy is a cosmological constant,
so that $\Omega_{\rm de}\equiv \Omega_\Lambda$, and $\alpha\equiv
3(1+w_{\rm de})$ is zero, since $w_{\rm de}\equiv w_\Lambda=-1$.
Using this distance measure, Riess et al. (2004) find that the ``gold
sample" of 157 SNe Ia is consistent with an $\Omega_{\rm m}=0.27$,
$\Omega_\Lambda=0.73$ cosmology, yielding $\chi_{\rm dof}^2=1.13$.
Adding several free parameters, specifically an acceleration
parameter $q_0\equiv -{\ddot{a}(t_0)a(t_0)}/{\dot{a}}(t_0)^2$ and $dq/dz$
evaluated at $z=0$, Riess et al. (2004)
find an even better fit with $\Omega_{\rm m} =0.3$ and
$\Omega_\Lambda=0.7$, yielding $\chi_{\rm dof}^2=1.06$.

At face value, this is a reasonable fit. The caveat, of course,
is that one must use many free parameters with this model.
One should also question the validity of introducing two new
parameters ($q_0$ and $dq/dz$) independent of $\Omega_{\rm m}$
and $\Omega_{de}$, given that the expansion history of the universe
in $\Lambda$CDM is completely specified once the latter two are
selected. As it turns out, the additional free parameters improve the
fit because the current acceleration needs to be counterbalanced
by an earlier deceleration that together yield an overall expansion
consistent with a coasting universe (i.e., $\langle q\rangle=0$,
equivalent to $\langle w\rangle=-1/3$).

In contrast, the luminosity distance in a universe with $R_{\rm h}=ct$
is given by the expression
\begin{equation}
d_{\rm L}=(1+z)R_{\rm  h}(t_0)\,\ln(1+z)
\end{equation}
(see also Melia 2009). Here, the only parameter is the Hubble
constant $H_0$, which enters through our gravitational radius
$R_{\rm h}(t_0)$. This is the proper form of the luminosity
distance to use in the analysis of Type Ia supernova data if
our understanding of the relativistic constraint $R_{\rm h}
=ct$ is correct. However, this form of the luminosity distance,
without the luxury of extra free parameters, does not fit the
current sample of Type Ia supernova as well as Equation~(32).

Interestingly, Equation~(33) {\it does} fit the data adequately
at low and high redshifts, but not in between, as may be seen, 
e.g., in figure~6 of Riess et al. (2004). This could be an important 
clue, because the difficulty with interpreting the data at intermediate 
redshifts is made more evident through a comparison of the gold 
sample with other, newer compilations. Though all of the currently 
available SNe Ia catalogs yield a consistent and robust value of
$\Omega_{\rm m}$ (i.e., $\approx 0.27$), they vary significantly
when it comes to the inferred redshift $z_{\rm acc}$ at which deceleration
is meant to have switched over to acceleration in the present epoc.
For example, the gold sample gives a value $z_{\rm acc}=0.46\pm 
0.13$ (Riess et al. 2004). The so-called Union2 sample contains 557 
events in the redshift range $0.015<z<1.4$ (Amanullah et al. 2010).
The analysis of these data alone yield $z_{\rm acc}\approx
0.75$, though with a fairly large uncertainty ($\pm 0.35$),
and a combination of the Union2 sample with the CMB
measurements yield $z_{\rm acc}=1.2\pm 0.10$. The ESSENCE
SNe Ia data span the redshift range $z=0.2-0.8$ (Wu et al.
2008). Their analysis yields a transition redshift $z_{\rm acc}
\approx 0.632$, roughly in the range of the others, but
not as tightly consistent with them as the value of $\Omega_{\rm m}$,
which ESSENCE finds to be $\approx 0.278$, quite close to the
value calculated from both the gold and Union2 samples.

\section{Discussion}
We draw several conclusions from this comparison. It is possible,
though we believe unlikely, that $\Lambda$CDM is correct
after all and that Equation~(27) is simply a coincidence, as
improbable as that may be. It would then be incumbent upon us
to understand where our argument for the constraint
$R_{\rm h}=ct$ has gone wrong. We stress, however, that we have
examined the need for this equality only for a flat cosmology
(i.e., $k=0$). The disparity between this condition and the Type
Ia supernova data may be telling us that the universe is not flat
after all---if it turns out that the constraint $R_{\rm h}=ct$
does not apply when $k\not=0$. We will examine this situation
next and report the results elsewhere.

On the other hand, it could very well be that $\Lambda$CDM is
currently providing a reasonable fit to the Type Ia supernova
data only because (i) it has several free parameters, some of them
($q_0$ and $dq/dz$) possibly inconsistent with the others
(e.g., $\Omega_{\rm m}$ and $\Omega_{\rm de}$); and (ii) other
factors, perhaps astrophysical in origin, are biasing the observed
supernova luminosities at intermediate redshifts. Certainly, the
fact that $z_{\rm acc}$ varies widely from sample to sample
could be an indication that this might be happening.

Of course, there are many other consequences of the $R_{\rm h}=ct$ 
constraint, e.g., with regard to baryogenesis, nucleosynthesis, and 
structure formation, all of which would have been affected in terms of 
when they could have occurred, if not the physical conditions prevalent 
at those times. Although it is beyond the scope of the present work to 
fully explore all of these processes, a detailed account is necessary 
before the viability of our proposal can be fully assessed. 

This extended analysis is necessary because the current situation 
with the standard model is far from adequate. For example, $\Lambda$CDM
does not provide a compelling explanation for the galaxy correlation 
function. Over the past four decades, the successively
larger galaxy redshift surveys have mapped the distribution of
galaxies with ever increasing precision, confirming correlation functions
consistent with a single power law on all scales (e.g., Marzke 
et al. 1995; Zehavi et al. 2002), from large regions
($r>10$ Mpc) exhibiting slight density fluctuations, to collapsed,
virialized galaxy groups and clusters ($r< 1$ Mpc). The lack of
any observational feature signaling the transition from one physical
domain to the next is surprising when viewed within context of
the standard model (see, e.g., Li \& White 2009), because
the matter correlation function in the concordance
model differs significantly from a power law. 

The most recent attempts at accounting for the unexpected galaxy
correlation function have relied on several new, fine-tuning additions
in order to get the correct profile (see, e.g., Watson et al. 2011). But the 
various contributing effects are intertwined and no simple, universal 
rule exists for which a power-law correlation function emerges.The
evolving competition between accretion and destruction rates of
subhalos over time is {\it required} to have struck just the right
balance at $z\approx 0$, leading Watson et al. (2011) to conclude 
that the power-law galaxy correlation function is a cosmic coincidence.

Part of the difficulty with this type of analysis is that, besides gravity
and pressure, other physical processes can play an important role
in the formation of structure, and these are not easy to handle. 
For example, in baryonic models, the most important physical
phenomenon is the interaction between baryons and photons during the
pre-recombination era, and the consequent dissipation due to viscosity
and heat conduction. 

Insofar as the $R_{\rm h}=ct$ universe is concerned, we can leave 
these elements aside for the moment, and at least suggest how the 
fundamental equation describing the dynamical growth of density 
fluctuations would appear in this cosmology.
Defining the density contrast $\delta\equiv \delta\rho/\rho$ in terms
of the density fluctuation $\delta \rho$ and unperturbed density $\rho$,
we can form the wavelike decomposition
\begin{equation}
\delta=\sum_\kappa \delta_\kappa(t)e^{i\vec{\kappa}\cdot{\bf r}}\;,
\end{equation}
where the Fourier component $\delta_\kappa$ depends only on cosmic time
$t$, and $\vec{\kappa}$ and {\bf r} are the co-moving wavevector and radius,
respectively. In the linear regime, the $\kappa$-th perturbative mode
satisfies the equation
\begin{equation}
\ddot{\delta}_\kappa +2{\dot{a}\over a}\delta_\kappa=\left({4\pi G\over c^2}\rho
-{v_s^2\kappa^2\over a^2}\right)\delta_\kappa\;,
\end{equation}
where a dot signifies differentiation with respect to $t$, $a=a(t)$ is
the cosmic expansion factor we defined earlier, and $v_s^2\equiv
dp/d\rho$ is the adiabatic sound speed squared, in terms of the
pressure $p$ and energy density $\rho$ (see, e.g., Tsagas 2002).

The second term on the left is due to the cosmic expansion and always
suppresses the growth of $\delta_\kappa$. The combined term on the right
reflects the conflict between gravity ($4\pi G\rho/c^2$) and pressure
support ($-{v_s^2\kappa^2/ a^2}$). Defining the proper wavelength of
the perturbation $\lambda\equiv 2\pi a/\kappa$, one sees immediately that
whether gravity or pressure support dominates depends on whether
$\lambda$ is greater or smaller than the so-called Jeans length
\begin{equation}
\lambda_J\equiv v_s\sqrt{\pi c^2\over G\rho}\;.
\end{equation}
In the standard model, one solves Equation~(35) by first
choosing the constituents of the universe (e.g., baryonic matter, cold
dark matter, and radiation) contributing to $\rho$, adopting an equation
of state to calculate $p$ and therefore $v_s$, and then integrating
$\delta_\kappa$ over time from an assumed set of initial conditions.

The origin of the initial seed perturbations is uncertain, one possible explanation
being that they are quantum fluctuations boosted to macroscopic scales by
inflation. The primordial power spectrum is usually assumed to have a power-law
dependence on scale,
\begin{equation}
P(\kappa)= A \kappa^n\;,
\end{equation}
with a scale-invariant spectral index $n=1$, and an unknown normalization factor $A$ 
that must be determined observationally. The initial conditions for the solution to 
Equation~(35) follow from this because at any redshift $z$, the power spectrum 
may also be written
\begin{equation}
P(\kappa,z)=\langle|\delta_\kappa(z)|^2\rangle\;,
\end{equation}
so the starting size of the fluctuation is
\begin{equation}
\delta_\kappa\propto \kappa^{1/2}\;.
\end{equation}

Equation~(35) is adequate for most applications, but not in situations where
the pressure is a significant fraction of $\rho$. In general relativity, both
$\rho$ and $p$ contribute to the ``active" mass inducing curvature, 
as evidenced by the appearance of both $\rho$ and $p$ in Equations~(4) and (5).
Thus, to analyze the growth of perturbations in an $R_{\rm h}=ct$ universe,
we must resort to the relativistic version of Equation~(35). Fortunately, this
transition is greatly simplified by the very simple equation of state implied by
the condition $R_{\rm h}=ct$, given by
\begin{equation}
p=w\rho
\end{equation}
with $w=-1/3$, as we discussed earlier.

For a universe with density $\rho$ and pressure $p=w\rho$, the linear
relativistic version of Equation~(35) is
\begin{equation}
\ddot{\delta}_\kappa+\left(2-6w+3v_s^2\right){\dot{a}\over a}\dot{\delta}_\kappa
-{3/2}\left(1+8w-3w^2-6v_s^2\right)\left({\dot{a}\over a}\right)^2\delta_\kappa=
-{\kappa^2 v_s^2\over a^2}\delta_\kappa\;.
\end{equation}
Therefore, for an $R_{\rm h}=ct$ universe, the dynamical equation for
$\delta_\kappa$ is
\begin{equation}
\ddot{\delta}_\kappa+{3\over t}\dot{\delta}_\kappa={1\over 3}c^2\left({\kappa\over a}\right)^2\delta_\kappa\;.
\end{equation}
We need to emphasize several important features of this equation. First of all, the active
mass in this universe is proportional to $\rho+3p=0$, and therefore the gravitational term
normally appearing in the standard model is absent (see Equation~35). But this does
not mean that $\delta_\kappa$ cannot grow. Instead, because $p<0$, the (usually dissipative)
pressure term in Equation~(35) here becomes an agent of growth. Moreover, there
is no Jeans length scale. In its place is the gravitational radius, which we can see most
easily by writing Equation~(42) in the form
\begin{equation}
\ddot{\delta}_\kappa+{3\over t}\dot{\delta}_\kappa-{1\over 3}{\Delta_\kappa^2\over t^2}\delta_\kappa=0\;,
\end{equation}
where 
\begin{equation}
\Delta_\kappa\equiv {2\pi R_{\rm h}\over \lambda}\;.
\end{equation}
Note, in  particular, that both  the gravitational radius $R_{\rm h}$ and the fluctuation
scale $\lambda$ vary with $t$ in exactly the same way, so $\Delta_\kappa$ is therefore
a constant in time. But the growth rate of $\delta_\kappa$ depends critically on whether
$\lambda$ is less than or greater than $R_{\rm h}$.

A simple solution to Equation~(43) is the power law
\begin{equation}
\delta_\kappa(t)=\delta_\kappa(0)t^\alpha\;,
\end{equation}
where evidently
\begin{equation}
\alpha^2+2\alpha-{1\over 3}\Delta_\kappa=0\;,
\end{equation}
so that
\begin{equation}
\alpha=-1\pm\sqrt{1+\Delta_\kappa^2/3}\;.
\end{equation}
Thus, for small fluctuations ($\lambda<<R_{\rm h}$),
\begin{equation}
\delta_\kappa\sim C_1 \kappa^{1/2}t^{\Delta_\kappa/\sqrt{3}}+C_2\kappa^{1/2}t^{-\Delta_\kappa/\sqrt{3}}\;,
\end{equation}
whereas for large fluctuations ($\lambda>>R_{\rm h}$),
\begin{equation}
\delta_\kappa\sim C_3\kappa^{1/2}+C_4\kappa^{1/2}t^{-2}\;,
\end{equation}
where the $C_i$ constants depend on the initial conditions.

Beyond this point there are too many unknowns to pin down  the final galaxy correlation 
function resulting from these growth functions. For example, we don't know how to set 
the values of $C_1$, $C_2$, $C_3$ and $C_4$ in a
model-independent way, nor does any of this analysis take into account the
non-linear growth that follows. But already we can point to a decided advantage
of the $R_{\rm h}=ct$ universe over $\Lambda$CDM. Whereas the concordance
model predicts different distributions at different scales, in part because of the influence
of the Jeans length, no such transition region exists for the $R_{\rm h}=ct$ universe. 
Instead, the fluctuation growth is driven by the pressure term, which looks the same no 
matter the perturbation length $\lambda<<R_{\rm h}$. At least in this regard, the
$R_{\rm h}=ct$ universe appears to be a better match to the observations.

\section{Conclusion}
Fortunately, a resolution to the $\Lambda$CDM versus $R_{\rm h}=ct$
universe dilemma will surely come with the observation
of standard candles at redshifts even greater than 1.8 (roughly
the current upper limit to the Type Ia samples). A cosmology 
with the time-independent constraint $R_{\rm h}=ct$
predicts a luminosity distance unmistakably distinguishable
from that of all other models. And the differences will manifest
themselves most prominently early in the universe's expansion
(i.e., at large redshift $z$), where all other models
(including $\Lambda$CDM) predict a rapid deceleration.

\section*{Acknowledgments}
This research was partially supported by ONR
grant N00014-09-C-0032 and an NSF Graduate Student Fellowship at
the University of Arizona, and by a Miegunyah Fellowship at the
University of Melbourne. FM is also grateful to Amherst College
for its support through a John Woodruff Simpson Lectureship.

\section*{Appendix}
The fact that $R_{\rm h}=ct$ in {\it both} an empty universe (Milne
1940) and a flat ($k=0$) universe is not a coincidence, as one may
appreciate from a simple heuristic argument justified by the
corollary to Birkhoff's theorem. As noted by Weinberg (1972),
the fact that the gravitational influence of any isotropic,
external mass-energy is zero within a spherical cavity, permits
the limited use of Newtonian mechanics to some cosmological
problems, which we can use here to gain some insight into the
dynamics implied by $k=0$.

Consider a sphere ``cut out" of a homogeneous and isotropic
universal medium with (proper) radius $R_{\rm s}(t)=a(t)r_{\rm s}$.
Adopting the Cosmological Principle, we assume that the density within
this region is a function of time $t$ only, and that every point within
and without the sphere expands away from every other point in proportion
to the time-dependent scale factor $a(t)$, which itself is the same
everywhere. According to Birkhoff's theorem and its corollary, we only
need to consider contributions to the energy from the contents enclosed
within $R_s$ to determine the local dynamics of this region extending
out to $R_s$.

Relative to an observer at the center of this sphere, the kinetic
energy of a shell with thickness $dR$ at radius $R$ is therefore
\begin{equation}
dK=4\pi R^2\,dR {1\over 2}{\rho(t)\over c^2}{\dot{R}}^2\;,
\end{equation}
and integrating this out from $r=0$ to $r=r_{\rm s}$, one easily
gets the total kinetic energy of this sphere relative to the
observer at the origin:
\begin{equation}
K={2\pi\over 5}{\rho(t)\over c^2}a^3{\dot{a}}^2\,{r_{\rm s}}^5\;.
\end{equation}

Let us now calculate the corresponding gravitational potential energy
of this spherical distribution (remember that this is a classical
approach). The potential energy of the shell at $R$ is
\begin{equation}
dV=-4\pi R^2\,dR\,{\rho(t)\over c^2}\,{GM(R)\over R}\;,
\end{equation}
where
\begin{equation}
M(R)={4\pi\over 3}\,R^3\,{\rho(t)\over c^2}
\end{equation}
is the total mass enclosed inside radius $R$. And integrating
this out from $r=0$ to $r=r_{\rm s}$, we see that the total
potential energy of this sphere (as measured by the observer
at the origin) is
\begin{equation}
V={16\pi^2G\over 15}{\rho(t)^2\over c^4}a^5{r_{\rm s}}^5\;.
\end{equation}

Classically, then, the observer measures a total energy of this
sphere given by
\begin{equation}
E={2\pi\over 5}{\rho(t)\over c^2}a^3{\dot{a}}^2\,{r_{\rm s}}^5-
{16\pi^2G\over 15}{\rho(t)^2\over c^4}a^5{r_{\rm s}}^5\;,
\end{equation}
which may be re-arranged to cast it into a more recognizable form:
\begin{equation}
\left({\dot{a}\over a}\right)^2={8\pi G\over 3c^2}\rho(t)+
{5c^2E\over 2\pi \rho(t)\,a^5\,{r_{\rm s}}^5}\;.
\end{equation}
Evidently, the local conservation of energy relative to the observer
at the origin is actually the Friedmann Equation (3), when we identify
the spatial curvature constant as
\begin{equation}
k\equiv -{10\over 3\,{r_{\rm s}}^2}\left({\epsilon\over\rho}\right)\;,
\end{equation}
where
\begin{equation}
\epsilon\equiv {3E\over 4\pi\,{R_{\rm s}}^3}
\end{equation}
is the {\it total} local energy density. A universe with positive curvature
therefore corresponds to a net negative energy, which means the system is
bound, whereas a negative curvature is associated with a positive total
energy density ($\epsilon>0$), characterizing an unbound universe.

A universe with net zero energy is therefore flat ($k=0$), and the
latest cosmological measurements (see, e.g., Spergel et al. 2003)
are apparently telling us that this is the state we're in. Let
us remember that general relativity is a local theory; it tells
us only about the gradient of the spacetime curvature locally due
to the presence of a source at that point. As far as general
relativity is concerned, therefore, the local dynamics of a universe
with net zero energy density ($\epsilon=0$) is indistinguishable from
an empty (or Milne) universe. This is the reason why $\ddot{a}=0$ in
both cases, and why $R_{\rm h}=ct$.

\end{document}